\documentclass[aps,letterpaper,prd,superscriptaddress,groupedaddress,preprintnumbers,floatfix,twocolumn,nofootinbib]{revtex4}
\usepackage{graphicx,comment}
\usepackage{dcolumn}
\usepackage{bm}
\usepackage{bbm}
\usepackage{amssymb}
\usepackage{array}
\usepackage[caption=false]{subfig}
\usepackage{multirow}
\usepackage{placeins}
\usepackage{epstopdf}
\usepackage{mathtools}
\usepackage{color}
\usepackage{cancel}
\usepackage{tikz,pgfplots,filecontents}
\pgfplotsset{compat=newest}
\usepackage{amsfonts}    
\usepackage{amsmath}
\usepackage{dcolumn}
\usepackage{algorithm}
\usepackage[compatible]{algpseudocode}

\renewcommand{\algorithmiccomment}[1]{\bgroup\hfill//~#1\egroup}
\newcommand{\paren}[1]{\left(#1\right)}
\newcommand{\curly}[1]{\left\{ #1 \right\}}
\newcommand{\netp}[0]{\tilde{p}_f}
\newcommand{\expt}[1]{\underset{#1}{\mathbb{E}}}
\newcommand{\pbp}[2]{\frac{\partial #1}{\partial #2}}
\newcommand{\sq}[1]{\left[ #1 \right]}
\newcommand{\avg}[1]{\left\langle #1 \right\rangle}

\newcommand{\obs}[0]{\mathcal{O}}

\DeclareMathOperator{\Exists}{\exists}

\begin{document}

\preprint{
\vbox{
\hbox{MIT-CTP/5114}
}}

\title{Flow-based generative models for Markov chain Monte Carlo in lattice field theory}

\author{M.~S.~Albergo}\affiliation{Perimeter Institute for Theoretical Physics, Waterloo, Ontario N2L 2Y5, Canada}\affiliation{Cavendish Laboratories, University of Cambridge, Cambridge CB3 0HE, U.K.}\affiliation{University of Waterloo, Waterloo, Ontario N2L 3G1, Canada}
\author{G.~Kanwar}\affiliation{Center for Theoretical Physics, Massachusetts Institute of Technology, Cambridge, MA 02139, U.S.A.}
\author{P.~E.~Shanahan}\affiliation{Center for Theoretical Physics, Massachusetts Institute of Technology, Cambridge, MA 02139, U.S.A.}
\affiliation{Perimeter Institute for Theoretical Physics, Waterloo, Ontario N2L 2Y5, Canada}
	
\begin{abstract}
A Markov chain update scheme using a machine-learned \emph{flow-based generative model} is proposed for Monte Carlo sampling in lattice field theories. The generative model may be optimized (trained) to produce samples from a distribution approximating the desired Boltzmann distribution determined by the lattice action of the theory being studied. Training the model systematically improves autocorrelation times in the Markov chain, even in regions of parameter space where standard Markov chain Monte Carlo algorithms exhibit critical slowing down in producing decorrelated updates. Moreover, the model may be trained without existing samples from the desired distribution. The algorithm is compared with HMC and local Metropolis sampling for $\phi^4$ theory in two dimensions.
\end{abstract}

\maketitle

\section{Introduction}

A key problem in lattice field theory and statistical mechanics is the evaluation of integrals over field configurations, referred to as path integrals. Typically, such integrals are evaluated via a Markov chain Monte Carlo (MCMC) approach: field configurations are sampled from the desired probability distribution, dictated by the action of the theory, using a Markov chain.
A significant practical concern is the existence of correlations between configurations in the chain. Critical slowing down~\cite{Wolff:1989wq} refers to the divergence of the associated autocorrelation time as a critical point in parameter space is approached. This behavior drastically increases the computational cost of simulations in these parameter regions~\cite{DelDebbio:2004xh,Meyer:2006ty}. For some models, algorithms have been found which significantly reduce or eliminate this slowing down~\cite{Kandel:1988zz,Bonati:2017woi,PhysRevD.98.054502,SwendsenWang87,Wolff1989,ProkofevSvistunov01,KawashimaHarada04,Bietenholz1995MeronCluster}, enabling efficient simulation. For field theories, a number of methods have been proposed to circumvent critical slowing down by variations of Hybrid Monte Carlo (HMC) techniques~\cite{Ramos:2012bb,Gambhir:2015rha,Cossu:2017eys,Jin2019QNHMC}, multi-scale updating procedures~\cite{Endres:2015yca,Detmold:2016rnh,Detmold:2018zgk}, open boundary conditions or non-orientable manifolds~\cite{Luscher:2012av,Mages:2015scv,Burnier:2017osu}, metadynamics~\cite{Laio:2015era}, and machine learning tools~\cite{Tanaka:2017niz,Shanahan:2018vcv}. In important classes of theories, however, critical slowing down remains limiting; for example, in lattice formulations of Quantum Chromodynamics (QCD, the piece of the Standard Model describing the strong nuclear force) it is a major barrier to simulations at the fine lattice spacings required for precise control of the continuum limit.

Here, a new \emph{flow-based MCMC} approach is proposed and is applied to lattice field generation. The resulting Markov chain has autocorrelation properties that are systematically improvable by an optimization (training) step before sampling.
In this method, samples $z$ are drawn from a simple distribution and then transformed by a change-of-variables (or ``flow'') $\phi = f^{-1}(z)$, resulting in samples $\phi$ with a new effective distribution $\netp$. The mapping $f^{-1}$ is chosen to be efficient to compute, making it easy to draw samples $\phi$, and is optimized within a variational family to produce a distribution $\netp$ close to the desired one.
To guarantee asymptotic exactness of sampling, a Markov chain is constructed using Metropolis-Hastings steps with $\tilde{p}_{f}$ taken as a proposal distribution. Since proposed samples are independent of the previous samples in the chain, the autocorrelation time and acceptance rate are coupled; the autocorrelation time drops to zero as the acceptance rate approaches $1$. This is true even in regions of parameter space where standard algorithms exhibit critical slowing down. Under mild conditions (detailed in Section~\ref{sec:flowMCMC}), this approach is guaranteed to generate samples from the desired probability distribution in the limit of a large number of updates.

This method has several features that make it attractive for the evaluation of path integrals in lattice field theories:
\begin{enumerate}
    \item The autocorrelation time of the Markov chain can be systematically decreased by training the model;
    \item
    Each step of the Markov chain requires only the model evaluation and an action computation;
    \item Each update proposal is independent of the previous sample, thus proposals can be generated in parallel and efficiently composed into a Markov chain;
    \item The model is trained using samples produced by the model itself, without the need for existing samples from the desired probability distribution. 
\end{enumerate}

Several other machine learning approaches have been applied to MCMC, for statistical mechanics systems, synthetic distributions, and simple lattice quantum field theories. Self-learning Monte Carlo (SLMC) methods have been applied fairly successfully to one- to three-dimensional Ising and fermionic systems. These methods construct, by a variety of techniques, an effective Hamiltonian for a theory that can be more easily sampled than the original Hamiltonian~\cite{Huang2017,Liu2017SLMC,Liu2016SLMCFermion, Nagai2017SLMCCT,Shen2018SLMCDNN}.
The effective Hamiltonian is learned using supervised learning techniques based on training data drawn from a combination of existing MCMC simulations, randomly-mutated samples, and the accelerated Markov chain itself (hence the term ``self-learning'').
Flow-based methods have been used for Monte Carlo sampling in the two-dimensional Ising model~\cite{Li2018}, many-body systems~\cite{NoeBoltzmannGenerators2018}, and
synthetic distributions~\cite{Song2017,Levy2017GenHMC}, and generative adversarial methods have been applied to two-dimensional scalar field theory~\cite{Urban2018,Zhou2018}.

In contrast to these approaches, the method proposed here focuses on directly generating samples from a close approximation to the true distribution in such a way that the exact likelihood of each produced sample is known. The direct generation allows self-learning as in SLMC, and the known likelihood allows use of a Metropolis-Hastings acceptance step to ensure exactness.

The proposed flow-based MCMC algorithm is detailed in Section~\ref{sec:flowMCMC}. A numerical study of its effectiveness in the context of two-dimensional $\phi^4$ theory is presented in Section~\ref{sec:phi4}. Finally, Section~\ref{sec:summary} outlines the further development and scaling of the approach that will be required for applications to theories defined in a larger number of spacetime dimensions and to more complicated field theories such as QCD.

\section{A flow-based Markov chain Monte Carlo algorithm}
\label{sec:flowMCMC}

\begin{figure*}
  \centering
  \begin{minipage}{.76\linewidth}
  \centering
  \subfloat[Normalizing flow between prior and output distributions]{
  \includegraphics[width=.97\linewidth]{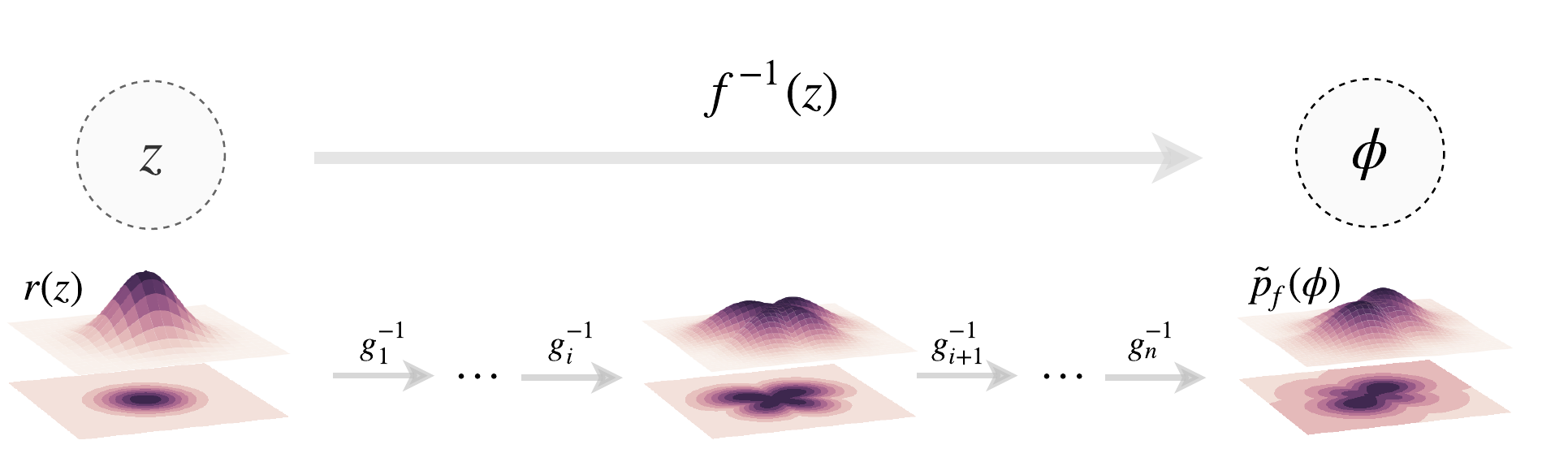}
  \label{subfl:norm-flow}
  }
  \end{minipage}%
  \begin{minipage}{.23\linewidth}
  \centering
  \subfloat[Inverse coupling layer]{
  \includegraphics[height=5.5cm]{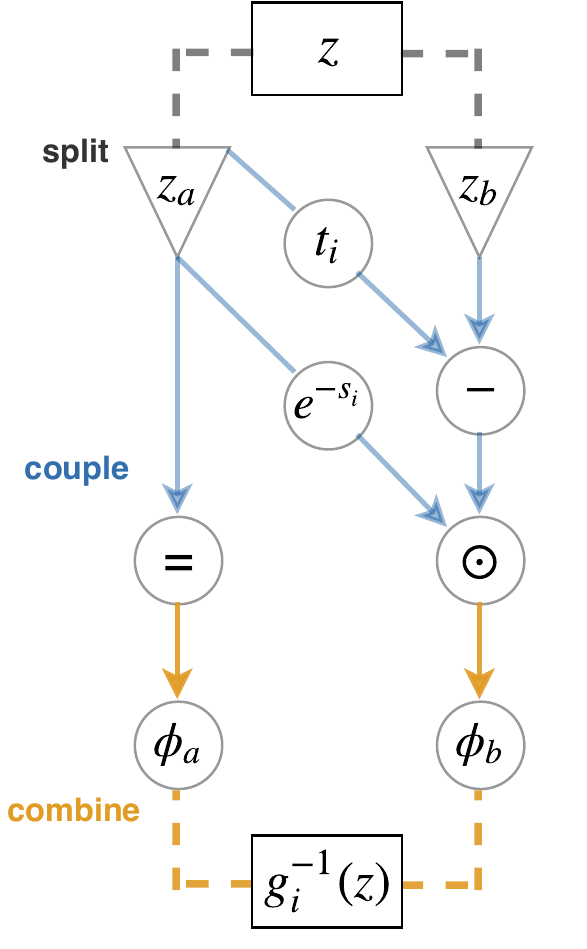}
  \label{subfl:inverse-coupling}
  }
  \end{minipage}
  \caption{ In \protect\subref{subfl:norm-flow}, a normalizing flow is shown transforming samples $z$ from a prior distribution $r(z)$ to samples $\phi$ distributed according to $\netp(\phi)$. The mapping $f^{-1}(z)$ is constructed by composing inverse coupling layers $g_i^{-1}$ as defined in Eq.~\eqref{eq:inv-affine-coupling} in terms of neural networks $s_i$ and $t_i$ and shown diagrammatically in \protect\subref{subfl:inverse-coupling}. By optimizing the neural networks within each coupling layer, $\netp(\phi)$ can be made to approximate a distribution of interest, $p(\phi)$.}
  \label{fig:flowdiag}
\end{figure*}

In lattice field theory, a Markov chain Monte Carlo (MCMC) process is an efficient way to generate field configurations $\phi \in \mathbb{R}^D$ distributed according to a target probability distribution
\begin{equation}
\begin{aligned}
    p(\phi) = e^{-S(\phi)} / Z, \quad
    \text{with}\quad Z =\int \prod_{j=1}^D d\phi_j \,e^{-S(\phi)},
\end{aligned}
\end{equation}
where $j$ indexes the $D$ components of $\phi$, $S(\phi)$ is the action that defines the theory and $Z$ is the partition function. Here, $\phi$ is defined to be a vector of $D$ real components representing the combined internal $(\alpha)$ and spacetime $(x)$ degrees of freedom of the field $\phi(x,\alpha)$ evaluated on a finite, discrete spacetime lattice (generalizations to gauge fields are discussed in Section~\ref{sec:summary}). A MCMC process generates a chain $\phi^{(0)} \rightarrow \phi^{(1)} \rightarrow \ldots \phi^{(N)}$ by steps through configuration space starting with an arbitrary configuration $\phi^{(0)}$. The steps are stochastic and are determined by the probabilities $T(\phi,\phi')$ associated with each possible transition $\phi \rightarrow \phi'$.
These probabilities must be non-negative and normalized:
\begin{equation}
    T(\phi,\phi')\ge 0
    \quad \text{and} \quad
    \int\prod_{j=1}^D d\phi'_j \, T(\phi,\phi') = 1.
\end{equation}
They must also satisfy the conditions of {\it ergodicity} and {\it balance} to ensure that samples in the chain are drawn from a distribution that converges to $p(\phi)$ after thermalization. For the chain to be ergodic, it must be possible to transition from a starting configuration $\phi$ to any other configuration $\phi'$ in a finite number of steps, i.e.,
\begin{equation}
    \Exists n \text{ such that } T^n(\phi,\phi')> 0 \text{ for all } \phi, \phi',
\end{equation}
and the chain must not have a period, for which it is sufficient that a single state has non-zero self-transition probability, i.e.,
\begin{equation}
    \Exists \phi \text{ such that } T(\phi,\phi) > 0.
\end{equation}
Balance is the condition that $p(\phi)$ is a stationary distribution of the transition:
\begin{equation}
    \int \prod_{j=1}^D d\phi_j \, p(\phi) T(\phi,\phi') = p(\phi').
\end{equation}
Any procedure which satisfies these conditions will, in the limit of a sufficiently long Markov chain, produce field configurations $\{\phi^{(i)}\}$ distributed according to $p(\phi)$. 

\subsection{Metropolis-Hastings with generative models}
\label{subsec:generative-metropolis}
Given a model that allows sampling from a known probability distribution $\tilde{p}(\phi)$, a Markov chain for a desired probability distribution $p(\phi)$ can be constructed via the independence Metropolis sampler, a specialization of the Metropolis-Hastings method~\cite{tierney1994}. For each step $i$ of the chain, an {\it update proposal} $\phi'$ is generated by sampling from $\tilde{p}(\phi)$, independent of the previous configuration. This proposal is accepted with probability  
\begin{equation}
A(\phi^{(i-1)},\phi') = \min\paren{1, \frac{\tilde{p}(\phi^{(i-1)}) }{ p(\phi^{(i-1)})} \frac{p(\phi')}{\tilde{p}(\phi')}}.
\label{eq:acc-rej}
\end{equation}
If the proposal is accepted, $\phi^{(i)} = \phi'$, otherwise $\phi^{(i)} = \phi^{(i-1)}$. This procedure defines the transition probabilities of the Markov chain.

The general Metropolis-Hastings algorithm has been proven to satisfy balance~\cite{hastings70} for any proposal scheme. For the independence Metropolis sampler, under the further condition that every state $\phi$ has non-zero proposal density and non-zero desired density,
\begin{equation}
    \tilde{p}(\phi) > 0, \; p(\phi) > 0 \text{ for all } \phi,
\end{equation}
the Markov chain is also ergodic and thus guaranteed to converge to the desired distribution~\cite{tierney1994}.

This Markov chain can be intuitively considered a method to correct an approximate distribution $\tilde{p}(\phi)$ to the desired distribution $p(\phi)$. The accept/reject statistics of the Metropolis-Hastings algorithm serve as a diagnostic for closeness of the approximate and desired distributions; if the distributions are equal, proposals are accepted with probability 1 and the Markov chain process is equivalent to a direct sampling of the desired distribution. This is made precise in Section~\ref{subsec:autocorr}.

\subsection{Sampling using normalizing flows}
\label{subsec:normalizing-flows}
Here, a {\it normalizing flow model} is used to define a proposal distribution $\tilde{p}(\phi)$ for a generative Metropolis-Hastings algorithm. Normalizing flows~\cite{rezende2015nf} are a machine learning approach to the task of sampling from complicated, intractable distributions. They do so by learning a map from an input distribution that is easy to sample to an output distribution that approximates the desired distribution. Normalizing flow models produce both samples and their associated probability densities, allowing the acceptance probability in Eq.~\eqref{eq:acc-rej} to be calculated.

A normalizing flow enacts the transformation between distributions by a change-of-variables\footnote{The convention of using $f^{-1}$ for the change-of-variables stems from typical applications of normalizing flows.}: a smooth, bijective function, $f^{-1}: \mathbb{R}^D \rightarrow \mathbb{R}^D$ maps samples $z$ from a prior distribution $r(z)$ to $\phi = f^{-1}(z)$. This mapping defines an output distribution $\netp(\phi)$, by the change-of-variables formula
\begin{align}
\label{eq:flowlikelihood}
\netp(\phi) = r(f(\phi))\left| \det  \frac{\partial f(\phi)}{\partial \phi} \right|.
\end{align}
Typically, the prior distribution is a simple and analytically-understood distribution (e.g., a normal distribution). While the desired distribution $p(\phi)$ is often complicated and difficult to sample from directly, optimizing the function $f$ allows one to generate samples from $\netp(\phi) \approx p(\phi)$. The function $f$ is chosen to have a tractable Jacobian such that the probability density $\netp(\phi)$ can be computed exactly according to Eq.~\eqref{eq:flowlikelihood}.
To encode a map from a simple distribution $r(z)$ to a complicated distribution $\netp(\phi)$, the map $f$ must be highly expressive while also being invertible and having a computable Jacobian. Here, the {\it real non-volume-preserving (real NVP) flow}~\cite{Dinh2016} machine learning approach is used: $f$ is constructed by the composition of {\it affine coupling layers} that scale and offset half of the components of the input at a time; the choice of which components of the data are transformed is part of the layer definition. Splitting the $D$-dimensional vector $\phi$ into ($D/2$)-dimensional pieces $\phi_a$ and $\phi_b$ according to this choice, a single coupling layer $g_i$ transforms $\phi$ to $z = g_i (\phi)$ via
\begin{align}\label{eq:affinecoupling}
    g_i(\phi) := \begin{cases}
    z_a = \phi_a  \\
    z_b = \phi_b \odot e^{s_i(\phi_a)} + t_i(\phi_a),
    \end{cases}
\end{align}
where $s_i$ and $t_i$ are neural networks mapping from $\mathbb{R}^{D/2}$ to $\mathbb{R}^{D/2}$ and $\odot$ denotes element-wise multiplication. Importantly, each layer $g_i$ is invertible without inverting the neural networks $s_i$ or $t_i$:
\begin{align}\label{eq:inv-affine-coupling}
    g_i^{-1}(z) := \begin{cases}
    \phi_a = z_a  \\
    \phi_b = (z_b - t_i(z_a)) \odot e^{-s_i(z_a)}.
    \end{cases}
\end{align}
The Jacobian matrix is lower-triangular and its determinant can be easily computed. For coupling layer $g_i$:
\begin{equation}
    \left| \det \pbp{g_i(\phi)}{\phi} \right| = \prod_{j=1}^{D/2} e^{\sq{s_i(\phi_a)}_j},
\end{equation}
where $j$ indexes the $D/2$ components of the output of $s_i$.
Stacking many coupling layers $g_1, \dots, g_n$ which alternate which half of the data is transformed, the function $f$ is defined as
\begin{equation}
    f(\phi) = g_1 (g_2 ( \dots g_{n} (\phi) \dots ) ).
\end{equation}
Using the chain rule, the determinant of the Jacobian of $f$ is a product of the contributions from each $g_i$. By increasing the number of coupling layers and the complexity of the networks $s_i$ and $t_i$, $f$ can systematically be made more expressive and general. Figure~\ref{fig:flowdiag} depicts how composing many coupling layers incrementally modifies a prior distribution which is easy to sample into a more complex output distribution that approximates a distribution of interest.

For a fixed initial distribution $r(z)$, the neural networks within each affine coupling layer of $f$ can be trained to bring $\netp(\phi)$ close to the desired distribution $p(\phi)$. This training is undertaken by minimizing a {\it loss function}. Here, the loss function used is a shifted Kullback-Leibler (KL) divergence\footnote{This training paradigm is a specific instance of Probability Density Distillation~\cite{Oord2017WaveNet, Li2018}.
} between the target distribution of the form $p(\phi) = e^{-S(\phi)}/Z$ and the proposal distribution $\netp(\phi)$:
\begin{equation}
\begin{aligned}\label{eq:loss}
L(\netp) &:= D_{KL}(\netp || p) - \log Z \\
&= \int \prod_j d\phi_j \, \netp(\phi) \paren{\log\netp(\phi) - \log p(\phi) - \log Z} \\
&=  \int \prod_j d\phi_j \, \netp(\phi) \paren{\log\netp(\phi) + S(\phi)}.
\end{aligned}
\end{equation}
This loss function has been successfully applied in related generative approaches to statistical lattice models~\cite{DBLP:journals/corr/abs-1809-10188,Li2018}. The formal shift by $\log{Z}$ in Eq.~\eqref{eq:loss} eliminates the need to compute the true partition function, and does not affect the gradients or location of the minima. By non-negativity of the KL divergence, the lower bound on the loss is $-\log{Z}$, and this minimum is achieved exactly when $\netp = p$.
In practice, the loss is stochastically estimated by drawing batches of $M$ samples from the model $\curly{\phi^{(i)} \sim \netp}$ and computing the sample mean:
\begin{equation}
\begin{aligned}\label{eq:loss-batch}
    \widehat{L\paren{\netp}} =
    \frac{1}{M} \sum_{i=1}^M \paren{\log \netp(\phi^{(i)}) + S(\phi^{(i)})}.
\end{aligned}
\end{equation}
The loss minimization can then be undertaken using stochastic optimization techniques such as stochastic gradient descent or momentum-based methods including Adam and Nesterov~\cite{Kingma2015Adam,nesterov1983}.

By construction, the flow model allows sampling from $\netp$ efficiently. The training process can thus be performed by drawing samples from the model itself, rather than using existing samples from the desired distribution as training data. This self-training is a key feature of the proposed approach to Monte Carlo sampling for field theories, where samples from the desired distribution are often computationally expensive to obtain. If samples do exist, they can be used to `pre-train' the network, although in the numerical studies undertaken here this was not found to be markedly more efficient in network optimization than using only self-training.

Given a trained model with distribution $\netp(\phi) \approx p(\phi)$, samples from $\netp$ can be used as proposals to advance a Markov chain using the generative Metropolis-Hastings algorithm described above. This forms the basis for the \emph{flow-based MCMC} algorithm proposed here:
\begin{enumerate}
    \item A flow-based generative model (here, a {real NVP} model) is trained using the shifted KL loss given in Eq.~\eqref{eq:loss} to have output distribution $\netp(\phi) \approx p(\phi)$;
    \item $N$ proposals $\curly{\phi'^{(i)} \sim \netp}$ are produced by sampling from the flow-based model (this can be done in parallel) and the associated action $S(\phi')$ is computed for each proposal;
    \item Starting from an arbitrary initial configuration, each proposed sample is successively accepted or rejected using the Metropolis-Hastings algorithm given in Eq.~\eqref{eq:acc-rej} to build a Markov chain of length $N$.
\end{enumerate}
When the prior distribution $r(z)$ is strictly positive, the invertibility and continuity of $f$ guarantees that the generated distribution $\netp(\phi)$ is also strictly positive. For all models with finite action, and thus $p(\phi) > 0$, the resulting Markov chain is then ergodic by the arguments detailed in Section~\ref{subsec:generative-metropolis}.

\subsection{Autocorrelation time for generative Metropolis-Hastings}
\label{subsec:autocorr}

For any Markov chain constructed via a generative Metropolis-Hastings algorithm (with independent update proposals), an observable-independent estimator for autocorrelation time can be defined from the accept/reject statistics of the chain. This serves both as a measure of the similarity between the proposal and desired distributions and enables proper error estimation for lattice observables~\cite{Wolff2003}.

Precisely, the autocorrelation at Markov chain separation $\tau$, for all observables, is given by the probability of $\tau$ rejections in a row,
\begin{equation}
p_{\tau\text{rej}} \equiv \avg{\prod_{i = 1}^\tau \mathbbm{1}_{\text{rej}}(i)} = \rho(\tau)/\rho(0),
\end{equation}
where $\mathbbm{1}_{\text{rej}}(i)$ is an indicator variable taking value $1$ when the proposed step from $i-1$ to $i$ was rejected in the Metropolis-Hastings algorithm, and $0$ otherwise. In practice, for a near-equilibrium, finite Markov chain with length $N$, a finite-sample estimator provides a good approximation to $\rho(\tau)/\rho(0)$:
\begin{equation}
\widehat{\rho(\tau)/\rho(0)}_{\text{acc}} = \frac{1}{N-\tau} \sum_{j=1}^{N-\tau} \prod_{i = 1}^\tau \mathbbm{1}_{\text{rej}}(i+j).
\end{equation}
This measure of autocorrelation is consistent with the usual definition; it is shown in Appendix~\ref{app:bounds} that the standard estimator for the autocorrelation of any given observable $\obs$,
\begin{equation}
\widehat{\rho(\tau)/\rho(0)}_\obs
=  \frac{\frac{1}{N-\tau} \sum_{i=0}^{N-\tau-1} (\obs_i - \bar{\obs})(\obs_{i+\tau}-\bar{\obs})}{\frac{1}{N} \sum_{i=0}^{N-1} \paren{\obs_i - \bar{\obs}}^2},
\end{equation}
also converges to $p_{\tau\text{rej}}$ in the limit $N\rightarrow \infty$.

The qualitative relation between acceptance rate and autocorrelations gives a convenient measure of the autocorrelation characteristics of a Markov chain. Precisely, the autocorrelation at distance $\tau$ can be bounded in terms of the average acceptance rate $a = 1 - \mathbb{E}\sq{p_\text{rej}}$:
\begin{equation}
\rho(\tau)/\rho(0) = \expt{\phi \sim p}\sq{p_{\text{rej}}^\tau(\phi)} \geq \paren{\expt{\phi \sim p}\sq{p_{\text{rej}}(\phi)}}^\tau = (1-a)^\tau.
\label{eq:mean-acc-bound}
\end{equation}
Increasing the acceptance rate of an independence Metropolis sampler is thus a necessary condition to reduce autocorrelations. Additionally, $a = 1$ exactly when the proposal and desired distributions are equal. In this case, $p_\text{rej}(\phi) = 0$ for each $\phi$, and there are no autocorrelations. While bringing $a$ close to $1$ does not provide an upper bound on autocorrelation, stochastically improving a loss function that measures distance between distributions is expected to reduce autocorrelations on average. In practice, autocorrelations should be evaluated as a test metric alongside the training loss to confirm improvement over the course of training the model. The correspondence between loss minimization, acceptance rate, and autocorrelations is studied in the context of $\phi^4$ theory in Section~\ref{sec:phi4}, where a clear correlation between $a$ and $\rho(\tau)/\rho(0)$ is observed.

\subsection{Critical slowing down}
\label{subsec:CSD-theory}

When a distribution is sampled using a Markov chain with large autocorrelation time, many updates are required to produce decorrelated samples. Critical slowing down (CSD) is defined as the divergence of the autocorrelation time of Markov chain sampling as a critical point in parameter space is approached~\cite{Wolff:1989wq}.
A numerically-stable definition of the characteristic autocorrelation time of a Markov chain is the \emph{integrated autocorrelation time}:
\begin{equation}
    \tau_\mathcal{O}^\text{int} = \frac{1}{2} + \lim_{\tau_\text{max} \rightarrow \infty}\sum_{\tau=1}^{\tau_\text{max}} \frac{\rho_\mathcal{O}(\tau)}{\rho_\mathcal{O}(0)}.
\end{equation}
As a critical point is approached, analysis of standard local-update algorithms for lattice models suggests $\tau_\mathcal{O}^\text{int}$ is typically well-described by a power law in the lattice spacing, or for fixed physical volume, a power law in the lattice sites per dimension, $L$. A dynamical critical exponent $z_\mathcal{O}$ is thus defined by a fit to $\tau_\mathcal{O}^\text{int} = \alpha_\mathcal{O} L^{z_\mathcal{O}}$ along a line of constant physics. An update algorithm for which the critical exponent is zero is unaffected by CSD.

In any generative Metropolis-Hastings simulation, the autocorrelation time is completely fixed by the expected accept/reject statistics, which in turn result from the structure of the proposal and desired distributions. For models trained with a target value of the integrated autocorrelation time used as a stopping criterion, CSD associated with the Markov chain sampling is thus trivially removed at the expense of up-front training costs. The difficulty of CSD is in essence shifted to the training of the model, i.e., to the optimization of the proposal distribution. The cost and scaling of this optimization task for $\phi^4$ theory, as studied here, and the prospects of scaling this approach to more complicated theories, are discussed in Sec.~\ref{subsec:training-cost}.

\section{Application of flow-based MCMC to $\phi^4$ theory}
\label{sec:phi4}

The theory with a massive scalar field $\phi(x)$ and a quartic self-interaction is one of the simplest interacting field theories that can be constructed. It is thus a convenient testing ground for new algorithms for lattice field theory, such as the flow-based MCMC approach proposed here. 

In a $d$-dimensional Euclidean spacetime, a discretized formulation of $\phi^4$ theory can be defined on a lattice with sites $x_\mu = a n_\mu$, where $a$ denotes the lattice spacing, $\mu \in \curly{1, \dots, d}$ labels spacetime dimension, and $n_\mu \in \mathbb{Z}^d$. Here, a finite lattice volume $V=(aL)^d$ is considered, with periodic boundary conditions in all dimensions. The lattice action (in units where $a=1$) can be expressed as
\begin{equation}\small
    S(\phi) = \sum_{x} \left( \sum_y\phi(x)\Box(x,y)\phi(y) + m^2\phi(x)^2 + \lambda \phi(x)^4\right),
\end{equation}
where the parameters $m^2$ and $\lambda$ are the bare mass squared and bare coupling, respectively, and the lattice d'Alembert operator is defined by
\begin{equation}
    \sum_y\Box(x,y)\phi(y) = \sum_\mu \left(2\phi(x) -\phi({x-\hat{\mu}}) - \phi({x+\hat{\mu}})\right).
\end{equation}
By taking expectation values over the distribution $p(\phi) = e^{-S(\phi)}/Z$, observables in the theory can be estimated.
%

\begin{table}
    \centering
    \begin{ruledtabular}
    \begin{tabular}{c|ccccc}
    & E1 & E2 & E3 & E4 & E5\\\hline
         $L$ & $6$ & $8$ & $10$ & $12$ & $14$ \\
         $m^2$ & $-4$ & $-4$ & $-4$ & $-4$ & $-4$ \\
         $\lambda$ & $6.975$ & $6.008$ & $5.550$ & $5.276$ & $5.113$ \\
         $m_p L$ & $3.96(3)$ & $3.97(5)$ & $4.00(4)$ & $3.96(5)$ & $4.03(6)$
    \end{tabular}
    \end{ruledtabular}
    \caption{Parameters $\{m^2,\lambda\}$ of $\phi^4$ theory on $L\times L$ lattices used for numerical study of the flow-based MCMC algorithm proposed here. The coupling constants $\lambda$ have been chosen to approximately maintain constant $m_pL$ as $L$ is varied.}
    \label{tab:lattice-params}
\end{table}

\begin{figure*}
    \centering
    \subfloat[Flow-based MCMC trained to 50\% mean acceptance.]{
    \includegraphics{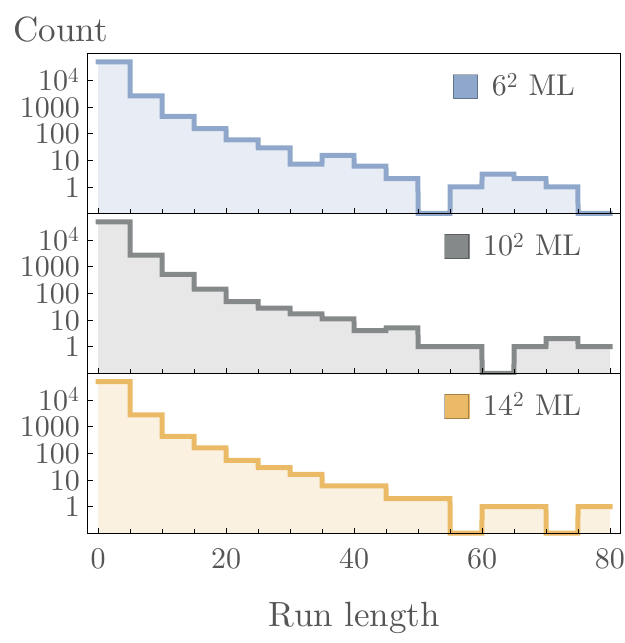}}
    \hspace{1cm}
    \subfloat[Flow-based MCMC trained to 70\% mean acceptance and HMC tuned to 70\% mean acceptance.]{
    \includegraphics{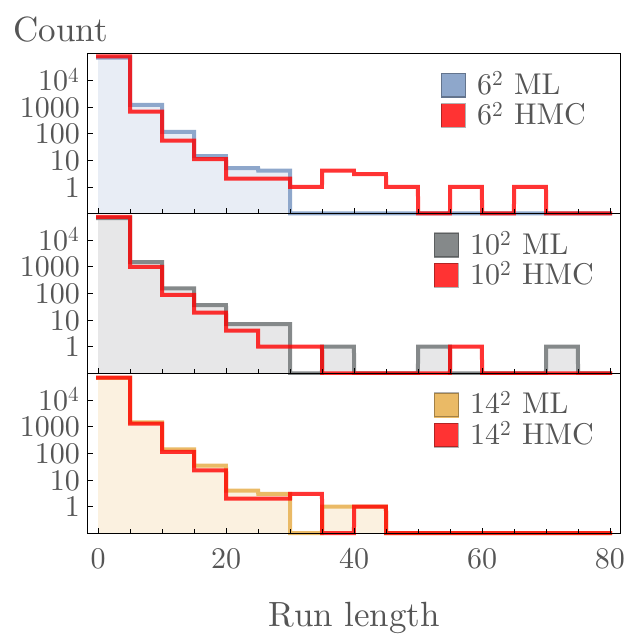}}
    \caption{Histograms of length of consecutive runs of Metropolis rejections in machine-learned (ML) models at both 50\% and 70\% mean acceptance. Also shown is the same statistic for Markov chains generated via HMC, where mean acceptance was tuned to 70\%. The frequency of long runs of rejections is consistently reduced for models trained to reach higher average acceptance. The ML and HMC ensembles at 70\% acceptance display very similar distributions of rejection streaks.}
    \label{fig:accepthistograms}
\end{figure*}

The observables studied here are the connected two-point Green's function
\begin{equation}
\begin{aligned}
    G_c(x) &= \frac{1}{V} \sum_y \big( \langle \phi(y)\phi(y+x) \rangle - \langle \phi(y) \rangle \langle \phi(y+x) \rangle \big)
\end{aligned}
\end{equation}
and its momentum-space representation
\begin{equation}
    \tilde{G}_c(\vec{p},t) = \frac{1}{L^{d-1}} \sum_{\vec{x}} e^{i\vec{p}\cdot\vec{x}}G_c(\vec{x},t),
\end{equation}
where $x_\mu = (\vec{x}, t)$,
as well as the corresponding pole mass
\begin{equation}
    m_p = -\partial_t \log \avg{\tilde{G}_c(0,t)},
\end{equation}
and the two-point susceptibility
\begin{equation}
    \chi_2 = \sum_x G_c(x).
\end{equation}
In the limit $\lambda \rightarrow \infty$, with $m^2/\lambda < 0$ fixed, scalar $\phi^4$ theory reduces to an Ising model.
Another observable of interest is therefore the average Ising energy density~\cite{vierhaus}, defined by
\begin{equation}
E = \frac{1}{d} \sum_{1\le \mu \le d} G_c(\hat{\mu}),
\end{equation}
where the sum runs over single-site displacements in all dimensions.

The action of $\phi^4$ theory is invariant under the discrete symmetry $\phi(x) \rightarrow -\phi(x)$. Depending on the value of the parameters $m^2$ and $\lambda$, this symmetry can be spontaneously broken. The theory thus has two phases: a symmetric phase and a broken-symmetry phase.

\subsection{Model definition and training}
\label{subsec:model}

For this proof-of-principle study, the flow-based MCMC algorithm detailed in Section~\ref{sec:flowMCMC} was applied to $\phi^4$ theory in two dimensions with $L=\{6,8,10,12,14\}$ lattice sites in each dimension. The parameters $m^2$ and $\lambda$ were chosen to fix  $m_p L \approx 4$ for each lattice size; their numerical values are given in Table~\ref{tab:lattice-params}. For simplicity in this initial work, all parameters were chosen to lie in the symmetric phase. In principle, the flow-based MCMC algorithm can be applied with identical methods to the broken-symmetry phase of the theory, but it remains to be shown that models can be trained for such choices of parameters.

\begin{figure*}
\begin{minipage}{0.48\textwidth}
    \centering
    \includegraphics{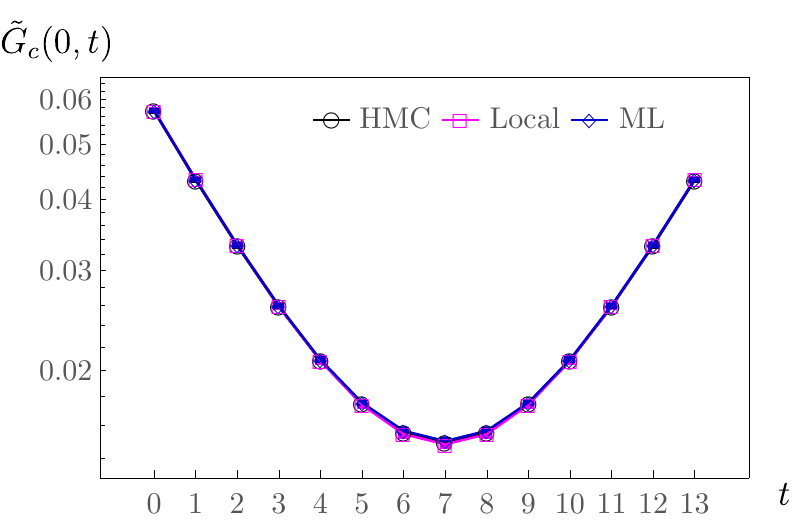}
    \caption{Zero-momentum Green's functions evaluated for parameter set E5. Results computed using $10^6$ configurations from the HMC, local Metropolis, and machine-learned (ML) ensembles are consistent within statistical errors. Error bars indicate 68\% confidence intervals estimated using bootstrap resampling with bins of size $100$.}
    \label{fig:observables-1}
\end{minipage}
\hspace{0.02\textwidth}
\begin{minipage}{0.48\textwidth}
    \centering
    \includegraphics{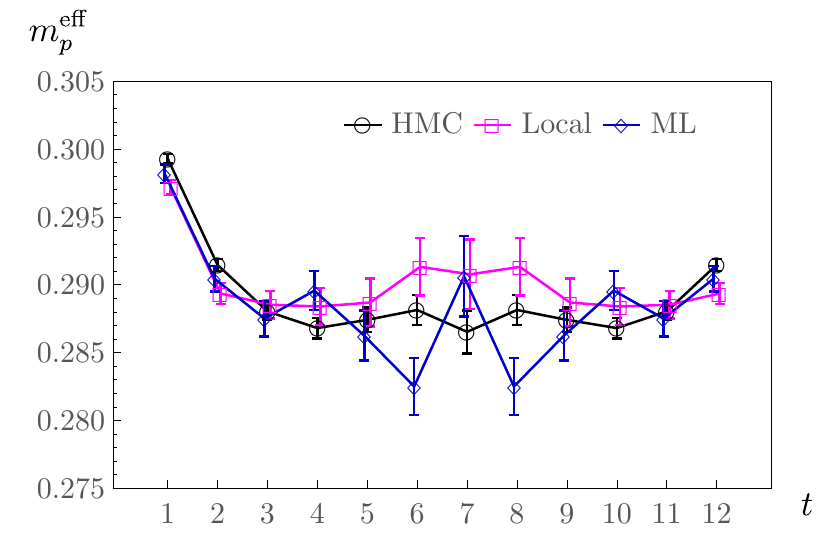}
    \caption{Effective pole masses evaluated for parameter set E5, defined by the arccosh estimator given in the main text. Results computed using $10^6$ configurations from the HMC, local Metropolis, and machine-learned (ML) ensembles are consistent within statistical errors. Error bars indicate 68\% confidence intervals estimated using bootstrap resampling with bins of size $100$.}
    \label{fig:observables-2}
\end{minipage}
\end{figure*}

For each set of parameters, real NVP models were defined using 8--12 affine coupling layers (see Sec.~\ref{subsec:normalizing-flows}). The coupling layers were defined to update half of the lattice sites in a checkerboard pattern; successive layers alternately updated the odd and even sites. The neural networks $s_i$ and $t_i$ used in coupling layer $g_i$ (see Eq.~\eqref{eq:affinecoupling}) were constructed from two to six fully-connected layers, each defined as multiplication by a rectangular matrix followed by pointwise application of a nonlinear function (here, a leaky rectified linear unit~\cite{Nair2010}). Intermediate vectors (hidden units) had sizes ranging between 100--1024. The prior distribution $r(z)$ was chosen to be an uncorrelated Gaussian distribution
\begin{equation}
    r(z) \propto \prod_i e^{-z_i^2 / 2}.
\end{equation}
The models were trained to minimize the shifted KL loss between the output distribution $\netp(\phi)$ and the desired distribution $p(\phi) = e^{-S(\phi)}/Z$ using gradient-based updates with the Adam optimizer~\cite{Kingma2015Adam}, a specific variety of gradient descent with momentum. A mean absolute error loss, defined in Appendix~\ref{app:mae-loss}, was optimized before training in the case of the $14^2$ model where it was found to accelerate convergence to the KL loss minimum.

An exhaustive study of the optimal choice of prior distribution $r(z)$, model depth, architecture and initialization of the neural networks, and of the mode of coupling of the affine layers, is beyond the scope of this proof-of-principle study. The parameters used here, however, proved to define sufficiently expressive models such that the Metropolis-Hastings algorithm applied to output from the trained models easily achieved acceptance rates of well over 50\%.
With further investment in hyperparameter optimization, higher rates of acceptance could be achieved. In any Markov chain using the Metropolis-Hastings algorithm, there is a tradeoff between computational cost and correlations resulting from low acceptance rates. The optimal acceptance rate minimizes the cost per decorrelated sample from the chain. Here, the cost of training, and not just model evaluation, must be considered, and the optimal level of training in future applications will depend on many factors, such as the desired ensemble size.

\begin{figure*}
\begin{minipage}{0.48\textwidth}
    \centering
    \includegraphics{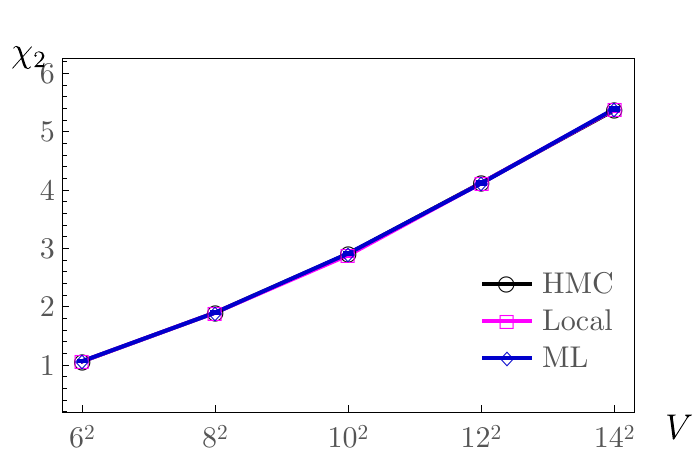}
    
    \includegraphics{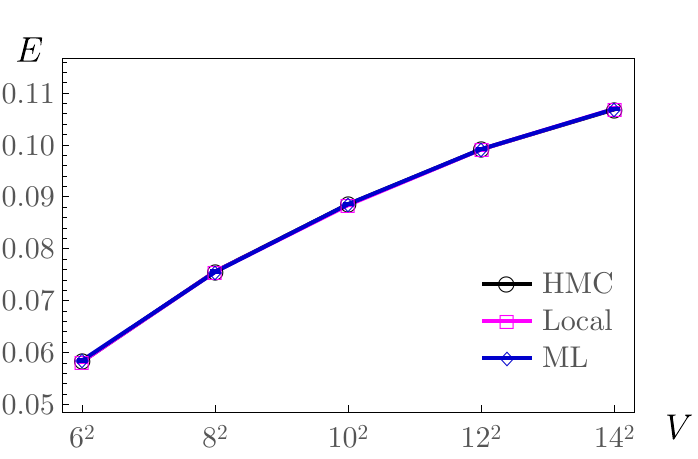}
    \caption{Susceptibility ($\chi_2$) and Ising energy ($E$) estimated on all ensembles. Results computed using $10^6$ configurations from the HMC, local Metropolis, and machine-learned (ML) ensembles are consistent within statistical errors. Errors indicate 68\% confidence intervals estimated using bootstrap resampling with bins of size $100$.}
    \label{fig:observables-3}
\end{minipage}
\hspace{0.02\textwidth}
\begin{minipage}{0.48\textwidth}
    \centering
    \includegraphics{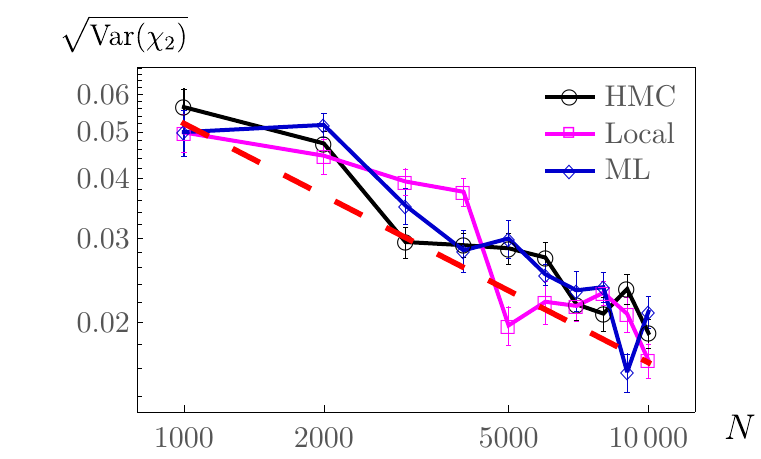}
    
    \includegraphics{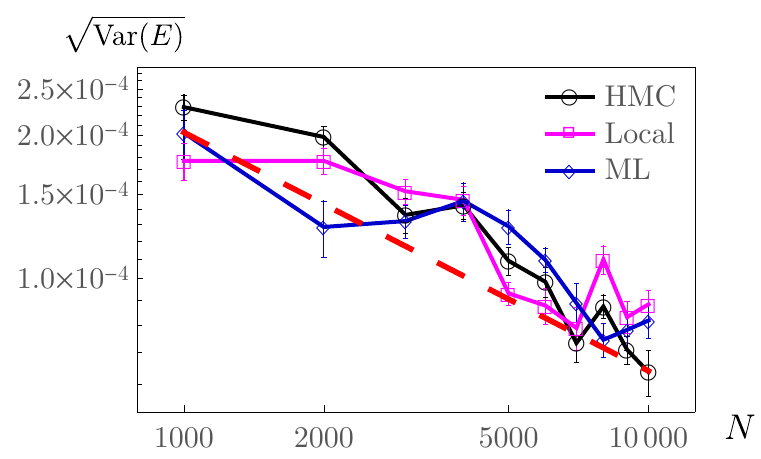}
    \caption{Statistical error varying with number of samples $N$ in two candidate observables, $\chi_2$ and $E$, for the HMC, local Metropolis, and machine-learned (ML) ensembles. The red dashed line shows a $1/\sqrt{N}$ curve normalized by the average error estimate of the three approaches at $N=1000$. Central values were estimated as 68\% confidence intervals on each observable by bootstrap resampling ensemble subsets of size $N$. Error bars indicate 68\% confidence intervals estimated using an external bootstrap resampling step.}
    \label{fig:errorscaling}
\end{minipage}
\end{figure*}

For each set of parameters studied, instances of the model were trained to reach both 50\% and 70\% average Metropolis acceptance. Figure~\ref{fig:accepthistograms} shows histograms of the number of updates between accepted configurations for models at both levels of training. Models trained to reach the higher acceptance rate are seen to have shorter runs of consecutive rejections.
Because autocorrelation is related to rejections by $\rho(\tau)/\rho(0) = p_{\tau \text{rej}}$ for independence Metropolis sampling, a reduced frequency of rejection runs with length longer than $\tau$ directly implies a reduction in $\rho(\tau)/\rho(0)$. Implications for critical slowing down of the generation of decorrelated configurations are discussed in Section~\ref{subsec:CSD}.

For comparison, ensembles of $10^6$ lattice configurations were generated using the machine-learned models in flow-based MCMC as well as standard local Metropolis~\cite{Wolff1996} and Hybrid Monte Carlo (HMC)~\cite{Duane1987HMC} algorithms at matched parameters. The local Metropolis algorithm employed a fixed order of sequential updates to each site, with proposed updates to $\phi(x)$ sampled uniformly from the interval $[\phi(x)-\delta, \phi(x)+\delta]$ followed by a Metropolis-Hastings accept/reject step; for all parameters considered, the width $\delta$ was tuned to achieve a 70\% acceptance rate. The HMC method was implemented using a leapfrog integrator with a fixed division of trajectory length $\tau$ into $10$ steps; the trajectory length $\tau$ was also tuned to achieve a 70\% acceptance rate. In both the local Metropolis and HMC methods, samples were saved after every 10th update.

\subsection{Tests: physical observables and error scaling}
\label{subsec:tests}

Since the flow-based MCMC algorithm satisfies ergodicity and balance, it is guaranteed to produce samples from the desired probability distribution in the limit of an infinite chain. To test the performance of the algorithm for a finite number of samples, each of the physical observables defined above was computed on ensembles of configurations at the parameters of Table~\ref{tab:lattice-params}, generated both using standard HMC and local Metropolis methods, as well as with the trained flow-based MCMC algorithm. Figures~\ref{fig:observables-1}--\ref{fig:observables-3} compare the observables computed on ensembles generated using all three methods.

To estimate the pole mass $m_p$, an effective mass is defined based on the zero-momentum Green's functions at various time separations:
\begin{equation}
    m^\text{eff}_p(t) = \text{arccosh}\paren{\frac{\tilde{G}_c(0,t-1) + \tilde{G}_c(0,t+1)}{2 \tilde{G}_c(0,t)}}.
\end{equation}
For all observables, the values computed using the flow-based MCMC ensembles are consistent within statistical uncertainties with those computed using the standard methods.
Moreover, Figure~\ref{fig:errorscaling} shows that the statistical uncertainties of the observables scale as $1 / \sqrt{N}$ with the number of samples $N$, as expected for decorrelated samples.

\subsection{Critical slowing down}
\label{subsec:CSD}

\begin{figure*}
    \subfloat[HMC ensembles]{
    \includegraphics{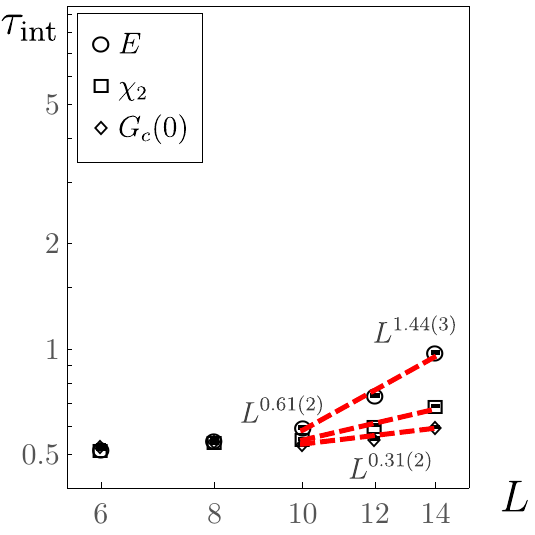}
    }%
    \subfloat[Local Metropolis ensembles]{
    \includegraphics{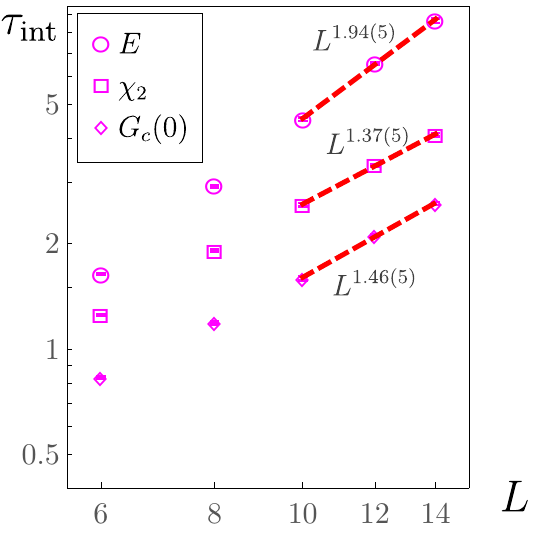}
    }%
    \subfloat[Flow-based MCMC ensembles]{
    \label{subfl:ml-auto}
    \includegraphics{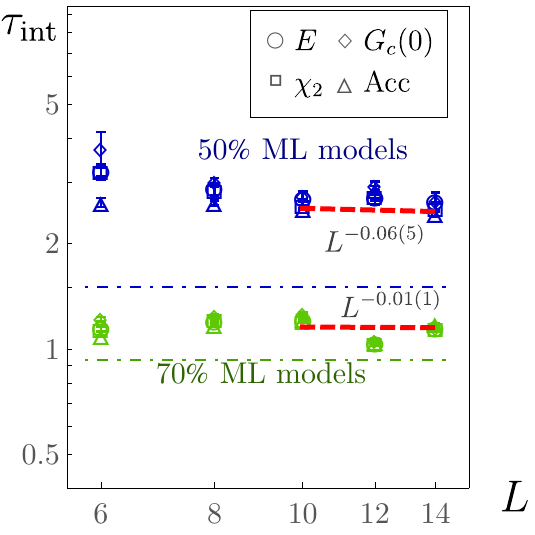}
    }
  \caption{
  Scaling of integrated autocorrelation time with respect to lattice size for HMC, local Metropolis, and flow-based MCMC. In \protect\subref{subfl:ml-auto} the upper sets of points in blue correspond to models trained to a mean acceptance rate of 50\%, while the lower sets of points in green correspond to models trained to a mean acceptance rate of 70\%. Dashed red lines display power law fits to $L = \curly{10,12,14}$ with labels $L^z$ specifying the scaling. The HMC and local Metropolis methods demonstrate power-law growth of $\tau_{\mathrm{int}}$, while $\tau_{\mathrm{int}}$ for the flow-based MCMC is consistent with a constant in $L$ and decreases as mean acceptance rate increases. Dot-dashed blue and green lines for the flow-based ensembles display lower bounds in terms of mean acceptance rate based on Eq.~\eqref{eq:mean-acc-bound}. Error bars indicate 68\% confidence intervals estimated by bootstrap resampling and error propagation.}
  \label{fig:compare_autocorr}

\end{figure*}

For $\phi^4$ theory, a number of algorithms have been developed that mitigate CSD to various extents, such as worm algorithms~\cite{vierhaus}, multigrid methods~\cite{GoodmanSokal89}, Fourier-accelerated Langevin updates~\cite{Batrouni87} and cluster updates via embedded Ising dynamics~\cite{BrowerTamayo89}. The path towards generalizing those algorithms to more complicated theories such as QCD, however, is not clear. Algorithms such as HMC and local Metropolis, which are also used for studies of QCD and pure gauge theory, exhibit CSD for $\phi^4$ (as well as more complicated theories) as the continuum limit is approached.

The parameter sets chosen for the study of $\phi^4$ theory in this work (Table~\ref{tab:lattice-params}) correspond to a critical line with constant $m_pL$ as $L\rightarrow \infty$. For the flow-based MCMC approach proposed here, as well as for ensembles generated using the HMC and local Metropolis algorithms, the autocorrelation times of the set of physical observables discussed previously were fit to leading-order power laws in $L$ to determine the dynamical critical exponents $z_\mathcal{O}$ for that observable.
Figure~\ref{fig:compare_autocorr} shows the autocorrelation times for each observable for each approach to ensemble generation. The absolute values of $\tau_\text{int}$ are not directly comparable between methods because the cost per update differs. The scaling with lattice size, on the other hand, indicates the sensitivity of each method to critical slowing down. For both HMC and local Metropolis, the critical behavior and consequently the performance of the algorithm was found to depend on the observable. In each case, the critical exponent was $0.3\lesssim z_\mathcal{O}\lesssim 2.0$. In comparison, for the flow-based MCMC ensembles at a fixed acceptance, the critical exponent was found to be consistent with zero, with the autocorrelation time observable-independent and in agreement with the acceptance-based estimator defined in Section~\ref{subsec:autocorr}.

Since the the mean acceptance rate was used as the stopping criterion for training these models, it was not guaranteed a priori that the measured integrated autocorrelation time would be constant across the different models used. The results in Figure~\ref{fig:compare_autocorr}, however, suggest that beyond the simple lower bound from Eq.~\eqref{eq:mean-acc-bound} there is a strong correlation between the mean acceptance rate and integrated autocorrelation time for models trained using a shifted KL loss. This is further confirmed by the similarity of the rejection run histograms across lattice sizes for flow-based MCMC, as shown in Figure~\ref{fig:accepthistograms}.

\subsection{Training costs}
\label{subsec:training-cost}
While CSD in the sampling step for the flow-based MCMC is eliminated, training the generative model introduces an additional up-front cost, as discussed in Section~\ref{subsec:CSD-theory}. Since this cost is amortized over the ensemble, this approach will naturally be computationally advantageous in the limit of generating a large number of samples. For a finite target ensemble size, the potential acceleration offered depends crucially on the training time.

In this work, all models were trained using one to two GPU-weeks, with the larger lattices incurring the most computational cost.
For the simple fully-connected architecture used in this work, the scaling of both the sampling and training time is controlled by dense matrix-vector multiplications which require $O(V^2)$ floating point operations each. The number of epochs used to train the largest lattice was also roughly $10\times$ that of the smallest lattice. This asymptotic scaling is a result of the simple model architecture used in this proof-of-principle study. For related methods applied  to image generation, using convolutional neural networks and a multi-scale architecture reduced training and sampling costs significantly and improved scaling to $O(V)$~\cite{Dinh2016}. There are physical grounds to expect these tools to apply equally well to the present application. Convolutional networks use only local information to update values in each layer, exploiting locality in the system, and use identical weights for each point on the lattice, manifestly preserving translational invariance. A multi-scale architecture learns coarse-grained distributions and fine-graining procedures in separate layers; this is an effective division of tasks for renormalizable quantum field theories, where simple coarse-grained descriptions are expected to arise. Generative models, and in particular flow-based models, are also rapidly evolving towards more efficient representation capacity. Complex coupling layers have been implemented~\cite{Dinh2015, Dinh2016}, as have generalized convolutions~\cite{Kingma2018, Hoogeboom2019} and transformations with continuous dynamics that are not dependent on restricted coupling layers~\cite{Grathwohl2019}. These developments allow models to better capture a distribution within a given number of training steps.

For complex applications, it is also critical that larger models with many coupling layers can be trained without exceeding memory bounds. The algorithm proposed here can be trained with constant memory cost as the number of layers is increased~\cite{Grathwohl2018}, alleviating the storage limitations that can arise in gradient-based optimization. Memory costs can be further reduced by distributing samples within each training batch across many machines.

Finally, typical applications seek to produce ensembles at many different choices of parameters, and often require parameter tuning. Training costs can therefore by amortized further; models trained with respect to an action at a given set of parameter values can either be used to initialize training or as a prior distribution for models targeting that action at nearby parameter values.

\section{Summary}
\label{sec:summary}

This work defines a flow-based MCMC algorithm to sample lattice field configurations from a desired probability distribution:
\begin{enumerate}
    \item A real NVP flow model is trained to produce approximately the desired distribution;
    \item Samples are proposed from the trained model;
    \item Starting from an arbitrary configuration, each proposal is accepted or rejected to advance a Markov chain using the Metropolis-Hastings algorithm.
\end{enumerate}
The approach is shown to define an ergodic and balanced Markov chain, thus guaranteeing convergence to the desired probability distribution in the limit of a long Markov chain. In essence, the flow-based MCMC algorithm combines the expressiveness of normalizing flows based on neural networks with the theoretical guarantees of Markov chains to create a trainable and asymptotically-correct sampler. Since these flows are applicable for arbitrary configurations with continuous, real-valued degrees of freedom, one can generically apply this method to any of a broad class of lattice theories. Here, the algorithm is implemented in practice for $\phi^4$ theory, and is demonstrated to produce ensembles of configurations that are indistinguishable from those generated using standard local Metropolis and HMC algorithms, based on studies of a number of physical observables.

A key feature of the approach is that models trained to a fixed acceptance rate do not experience critical slowing down in the sampling stage. In particular, the autocorrelation time for all observables is dictated entirely by the accuracy with which the flow model has been trained; perfect training corresponds to decorrelated samples and 100\% acceptance in the Metropolis-Hastings step of the MCMC process. Nevertheless, the efficiency with which the training step of this approach can be scaled to larger model sizes, and to more complicated theories such as QCD, remains to be studied. Recent advances in the training and scaling of flow models provide reasons for optimism on this front. Further, incorporating symmetries generally improves data efficiency of training, and implementing spacetime and gauge symmetries~\cite{Cohen2019GaugeEquiv} may be a natural next step to practically train these flow models for lattice gauge theories like QCD.

In moving towards lattice gauge theories such as QCD, several theoretical developments are also required. The real NVP model chosen to parameterize the normalizing flows here is described in terms of vectors of variables $\phi \in \mathbb{R}^D$. Gauge configurations, however, live in a compact manifold arising from the Lie group structure. Extending this method will require a normalizing flow model that can act on this manifold while remaining sufficiently expressive. The choice of prior likewise will need to be extended to a distribution over the manifold of lattice gauge configurations which can be easily sampled. A uniform distribution, for example, may be a candidate for a prior, but this choice must be tested in the context of a specific flow model.

If the flow-based MCMC algorithm proposed here can be implemented for a complex theory such as QCD, the advantages would be significant; arbitrarily large ensembles of field configurations could be generated at minimal cost. The independence of the proposal step from any previous configuration allows parallel generation of proposals, and the continually-improving support in hardware and software for neural network execution suggests future practical gains for this style of ensemble-generation. Given efficient sample generation from a trained model, ensembles would not need to be stored long-term. Moreover, a model trained for one action could either be re-trained or used as a prior for another flow model targeting an action with nearby parameter values. This would allow efficient tuning of parameters and generation of additional ensembles interpolating between and extrapolating from existing models.

\section*{Acknowledgements}

We thank J.-W. Chen, K.~Cranmer, W.~Detmold, R.~Melko, D.~Murphy, A.~Pochinsky, and B.~Trippe for helpful discussions.
This work is supported in part by the U.S.~Department of Energy, Office of Science, Office of Nuclear Physics under grant Contract Number DE-SC0011090. This research used resources of the Argonne Leadership Computing Facility, which is a DOE Office of Science User Facility supported under Contract DE-AC02-06CH11357, under the ALCF Aurora Early Science Program. Some work was undertaken at the Kavli Institute for Theoretical Physics, supported by the National Science Foundation under Grant No.~NSF PHY-1748958.
PES is supported by the National Science Foundation under CAREER Award 1841699, GK is supported by the U.S.~Department of Energy under the SciDAC4 award DE-SC0018121, and PES and MSA are supported in part by NSERC and the Perimeter Institute for Theoretical Physics. Research at the Perimeter Institute is supported by the Government of Canada through the Department of Innovation, Science and Economic Development and by the Province of Ontario through the Ministry of Research and Innovation.

\appendix

\section{Acceptance rate estimator for autocorrelation}
\label{app:bounds}

Here it is shown that the standard estimator for the autocorrelation of an observable $\mathcal{O}$ converges in the limit of infinite path length to  $p_{\tau \mathrm{rej}}$, as claimed in Section~\ref{subsec:autocorr}. The standard estimator is defined by:
\begin{align}
& \lim_{N\rightarrow\infty} \expt{}\sq{\widehat{\rho(\tau)/\rho(0)}_O} \\
=\,&
    \lim_{N\rightarrow\infty} \expt{} \sq{\frac{\frac{1}{N-\tau} \sum_{i=0}^{N-\tau-1} (\obs_i - \bar{\obs})(\obs_{i+\tau}-\bar{\obs})}{\frac{1}{N} \sum_{i=0}^{N-1} \paren{\obs_i - \bar{\obs}}^2}} \\
=\,& \lim_{N\rightarrow\infty} \expt{} \sq{\frac{(\obs_i - \bar{\obs})(\obs_{i+\tau}-\bar{\obs})}{\frac{1}{N} \sum_{i=0}^{N-1} \paren{\obs_i - \bar{\obs}}^2}},
\end{align}
where the final equality is true assuming the Markov chain is initialized with a sample from the stationary distribution $p(\phi)$ (i.e., it is assumed that enough prior iterations were discarded that the chain is thermalized). The expectation value can then be split by cases and, conditioning on the fixed accept/reject pattern, the expectation values can be computed by identifying the distributions of observables $\mathcal{O}_i$ and $\mathcal{O}_{i+\tau}$:
\begin{widetext}
\begin{align}
\lim_{N\rightarrow\infty} \expt{}\sq{\widehat{\rho(\tau)/\rho(0)}_O}
&= \lim_{N\rightarrow \infty} p_{\tau\text{rej}} \expt{} \sq{\frac{(\obs_i - \bar{\obs})(\obs_{i+\tau}-\bar{\obs})}{\frac{1}{N} \sum_{i=0}^{N-1} \paren{\obs_i - \bar{\obs}}^2}
    \Bigg| \text{all proposals } i+1, \dots, i+\tau \text{ rejected}} + \\
    &\qquad\qquad (1-p_{\tau\text{rej}}) \expt{} \sq{\frac{(\obs_i - \bar{\obs})(\obs_{i+\tau}-\bar{\obs})}{\frac{1}{N} \sum_{i=0}^{N-1} \paren{\obs_i - \bar{\obs}}^2}
    \Bigg| \text{some proposal } i+1, \dots, i+\tau \text{ accepted}} \\
&= p_{\tau\text{rej}} \lim_{N\rightarrow\infty} \expt{\phi\sim p} \sq{\frac{(\obs[\phi] - \bar{\obs})^2}{\frac{1}{N} \sum_{j \notin [i,i+\tau]} (\obs_j - \bar{\obs})^2 + O(\tau/N)}} + \\
    &\qquad\qquad (1-p_{\tau\text{rej}}) \lim_{N\rightarrow\infty} \expt{\phi\sim p} \expt{\phi' \sim \tilde{p}} \sq{ \frac{(\obs[\phi] - \bar{\obs})(\obs[\phi'] - \bar{\obs})}{\frac{1}{N} \sum_{j \notin [i,i+\tau]} (\obs_j - \bar{\obs})^2 + O(\tau/N)}} \\
&= p_{\tau\text{rej}} \frac{\rho(0)}{\rho(0)} + (1-p_{\tau\text{rej}}) \frac{\cancel{\expt{\phi\sim p} \sq{\obs[\phi] - \bar{\obs}}} \expt{\phi'\sim \tilde{p}} \sq{\obs[\phi'] - \bar{\obs}}}{\rho(0)} = p_{\tau\text{rej}}.
\end{align}
\end{widetext}
The limit $N \rightarrow \infty$ is used to drop biases arising from conditioning on behavior within the region $[i,i+\tau]$.

\section{Mean absolute error loss}
\label{app:mae-loss}
The mean absolute error (MAE) loss optimized before training some models is defined by:
\begin{equation}
L_\text{MAE}(\netp) := \int \prod_j d\phi_j \, \netp(\phi) \left| \log{\netp(\phi)} - \log{p(\phi)} \right|.
\end{equation}
It is bounded below by the KL divergence $D_{KL}(\netp || p)$ and has global minima exactly where the KL loss does, when $\netp = p$.

In practice, the loss is stochastically estimated by drawing batches of $M$ samples from the model $\curly{\phi^{(i)} \sim \netp}$ and computing the sample mean:
\begin{equation}
\begin{aligned}
\widehat{L_\text{MAE}(\netp)} &= \frac{1}{M} \sum_{i=1}^M \left| \log{\netp(\phi^{(i)})} - \log{p(\phi^{(i)})} \right| \\
&= \frac{1}{M} \sum_{i=1}^M \left| \log{\netp(\phi^{(i)})} + S(\phi^{(i)}) + \log{Z} \right|.
\end{aligned}
\end{equation}
To employ this loss, the partition function must either be estimated ahead of time, or initialized as a trainable parameter. In this study, a multistage method~\cite{ValleauFreeEnergy72} was used to estimate and fix the partition function value used while optimizing $L_\text{MAE}$.

\setlength{\parskip}{2pt}
This loss is appealing due to the point-by-point potential driving the distribution towards the correct one. Any errors in computing $\log{Z}$, however, result in a minimum at which the model distribution $\netp(\phi)$ does not necessarily agree with the desired distribution $p(\phi)$. This loss was therefore only used prior to training with the shifted KL loss.

\bibliography{FlowGenBib}

\end{document}